\documentclass[a4paper]{article}

\usepackage{spconf}
\usepackage{cite}
\usepackage{multirow}
\usepackage{tablefootnote}
\usepackage{footmisc}
\usepackage{pgfplots}
\usepackage{booktabs}

\title{Bandwidth Embeddings for Mixed-bandwidth Speech Recognition}
\name{Gautam Mantena, Ozlem Kalinli, Ossama Abdel-Hamid, Don McAllaster}
\address{
  Apple Inc., USA\\
  \small \{gmantena,~okalinli,~oabdelhamid,~dmcallaster\}@apple.com}

\begin{document}

\maketitle
\begin{abstract}
%Use of deep neural networks (DNNs) for acoustic modeling (AM) has driven the need for large amounts of training data. Ideally, one would like to use all available data for all tasks/domains; however multi-style training is not trivial and requires lots of tuning for how to mix the data. On the other hand, building a specific AM for each task/domain causes data fragmentation and limits the performance. 
In this paper, we tackle the problem of handling narrowband and wideband speech by building a single acoustic model (AM), also called mixed bandwidth AM. In the proposed approach, an auxiliary input feature is used to provide the bandwidth information to the model, and bandwidth embeddings are jointly learned as part of acoustic model training. Experimental evaluations show that using bandwidth embeddings helps the model to handle the variability of the narrow and wideband speech, and makes it possible to train a mixed-bandwidth AM. Furthermore,  we propose to use parallel convolutional layers to handle the mismatch between the narrow and wideband speech better, where separate convolution layers are used for each type of input speech signal. Our best system achieves $13\%$ relative improvement on narrowband speech, while not degrading on wideband speech.
\end{abstract}

\section{Introduction}
\label{sec:intro}
Currently, there are many devices and equipment that receive both narrowband  and wideband speech for automatic speech recognition (ASR) based applications.  In conventional approaches, different acoustic models (AMs) are built to handle narrow and wideband speech separately since their sampling frequencies are different (8 kHz vs 16 kHz). However, it is not very economical or efficient to collect large amounts of training data for each of the tasks. A simple solution is to downsample the wideband speech and treat it similar to that of the narrowband. However, wideband has information that is useful to detect certain phonemes and is lost with downsampling\cite{ICASSP:STERN:1994, SLT:LI:2012}. Moreover, models built on narrowband tends to perform worse on the wideband speech~\cite{TASLP:ALEX:2007, ICASSP:STERN:1994}. Hence, it is not a trivial task to build mixed-bandwidth AMs.

In general, bandwidth expansion (BWE) of speech can be used to convert narrowband to wideband speech \cite{INTERSPEECH:ALEX:2005, ICASSP:LEE:2015, INTERSPEECH:WANG:2015, INTERSPEECH:CHLEE:2015}. 
 BWE is a technique used to reconstruct the high frequency components of the narrowband using the correlation that exists between low and high frequency of the speech signal~\cite{INTERSPEECH:WANG:2015}. In \cite{ICASSP:LEE:2015, INTERSPEECH:WANG:2015, INTERSPEECH:CHLEE:2015}, deep neural network architectures such as feed forward network and a variant of restricted Boltzman machine (RBM) have been used to generate the higher frequency components. In \cite{SLT:LI:2012}, some issues reported for BWE are: (a) BWE is quite complicated and often introduces errors, and (b) in certain cases, the improvements in the recognition are seen only for less amounts of wideband speech ($\leq 50$ hrs of transcribed data). Thus modeling based approaches have been explored.

In \cite{TASLP:ALEX:2007, SLT:LI:2012}, mixed-bandwidth AM training is considered as a missing feature problem. That is, for narrowband speech, the spectral features represent information only from 0-4 kHz and the remaining 4-8 kHz are missing. 
%Hence, zero padding is applied to the features to compensate for the the missing spectral information. 
In \cite{SLT:LI:2012}, 22 and 27 filter bank filters were used for 8 kHz and 16 kHz data. The 22 filter bank features of 8 kHz data correspond to the 0-4 kHz of the 16 kHz data. To make sure all the features have the same dimension, zero padding is applied to the remaining 5 missing dimensions of the 8 kHz data. This approach is a simple and effective method. In \cite{TASLP:ALEX:2007}, training a Gaussian mixture model hidden Markov model (GMM-HMM) using a modified expectation maximization (EM) algorithm was proposed to handle these extended narrowband features. In recent years, deep learning based acoustic modeling have been shown to be successful for many state-of-the-art  automatic speech recognition (ASR) systems~\cite{SPM:HINTON:2012, TASLP:LIDENG:2012, ASRBOOK:DONG:2014}. Use of the extended features in combination with powerful AMs such as deep neural networks (DNNs) has shown to perform well on mixed-bandwidth speech~\cite{SLT:LI:2012, INTERSPEECH:YOU:2014}. In~\cite{SLT:LI:2012}, it has been shown that: (a) DNNs can learn the variations in the narrow and wideband speech, (b) a single DNN can be used to recognize mixed-bandwith speech, and (c) improved recognition performance can be achieved on wideband speech. It is important to note that, in these techniques, different feature extraction techniques are used for narrow and wideband speech. In this paper, we show that deep neural network based AMs are powerful and can handle such variations in the data automatically with the help of bandwidth embeddings. 
%Other modeling approaches as mentioned in 
Another modeling based approach was proposed in \cite{INTERSPEECH:XIAODAN:2017}, where narrowband data was limited and thus transfer learning was used to improve the performance of the system for the narrowband speech. There, a separate model was built for each of the narrow and wideband speech tasks.  Our work focuses on building a single model which performs well on both wide and narrowband speech and thus different from the work described in \cite{INTERSPEECH:XIAODAN:2017}.

%In this paper, we show that DNNs are powerful enough to handle this variability and there is no need for an explicit changes at the front-end. 

In this paper, we focus on a modeling approach for mixed-bandwidth speech recognition. AMs often tend to perform poorly on unseen data such as new speaker, different noise conditions, etc. To overcome these problems, techniques such as speaker or noise aware training have been explored~\cite{ICASSP:JAING:2013, ICASSP:JIANG:2014, ASRU:PICHENY:2013, ICASSP:SIM:2015, ICASSP:MANTENA:2016}. In these techniques, auxiliary information such as speaker codes~\cite{ICASSP:JAING:2013, ICASSP:JIANG:2014}, i-vectors \cite{ASRU:PICHENY:2013}, and bottleneck (BN) vectors \cite{ICASSP:SIM:2015, ICASSP:MANTENA:2016} are explicitly provided as input to the model. In \cite{CORR:KIM:2018, INTERSPEECH:SNYDER:2017}, speaker and noise embedding vectors are obtained by training another neural network classifier. 
%Unlike speaker and noise features, obtaining bandwidth related features is not a trivial task. Hence, 
Here, we propose to use an auxiliary input feature to the model indicating the bandwidth of the input speech, and bandwidth embeddings are jointly learned as part of the acoustic model training. 
%Hence, we employ the use of embedding to learn these auxiliary features as a part of the model training. 
This work is similar to the approach described in \cite{ICASSP:JAING:2013, ICASSP:JIANG:2014} where speaker representations  (also referred to as speaker codes) are learned during model training. To the best of our knowledge, there is no prior work to use embeddings for mixed-bandwith ASR. The major contributions of this work are as follows:
\begin{itemize}
	\item Use of embeddings to learn representations for narrow and wideband speech. We show that these features derived from the embedding layer can be used to capture the variations in the data and thus help us to build a mixed-bandwidth speech AM. The embedding vectors for narrowband and wideband speech are learned during the model training; hence easy to use. 
	\item We extend the use of bandwidth embeddings to a new model architecture which uses parallel convolutional layers to process narrow and wideband speech separately. 
	\item Experimental results show that we can build a single mixed-bandwidth model, and achieve a relative improvement of $13\%$ on upsampled narrowband speech in word error rate (WER) over the baseline system, while not degrading performance on the wideband speech.
\end{itemize}

The outline of the paper is as follows: Section~\ref{sec:embedded} describes the approach to learn bandwidth embeddings. In Section~\ref{sec:parallel_conv}, we describe the use of parallel convolutional layers for processing narrow and wideband speech separately. Section~\ref{sec:database} describes the database used and followed by detailed evaluations in Section~\ref{sec:evals}. Conclusions are provided in Section~\ref{sec:conclusions}.

\section{Bandwidth Embeddings and AM Training}
\label{sec:embedded}
\begin{figure}[t]
  \centering
  \includegraphics[width=\linewidth]{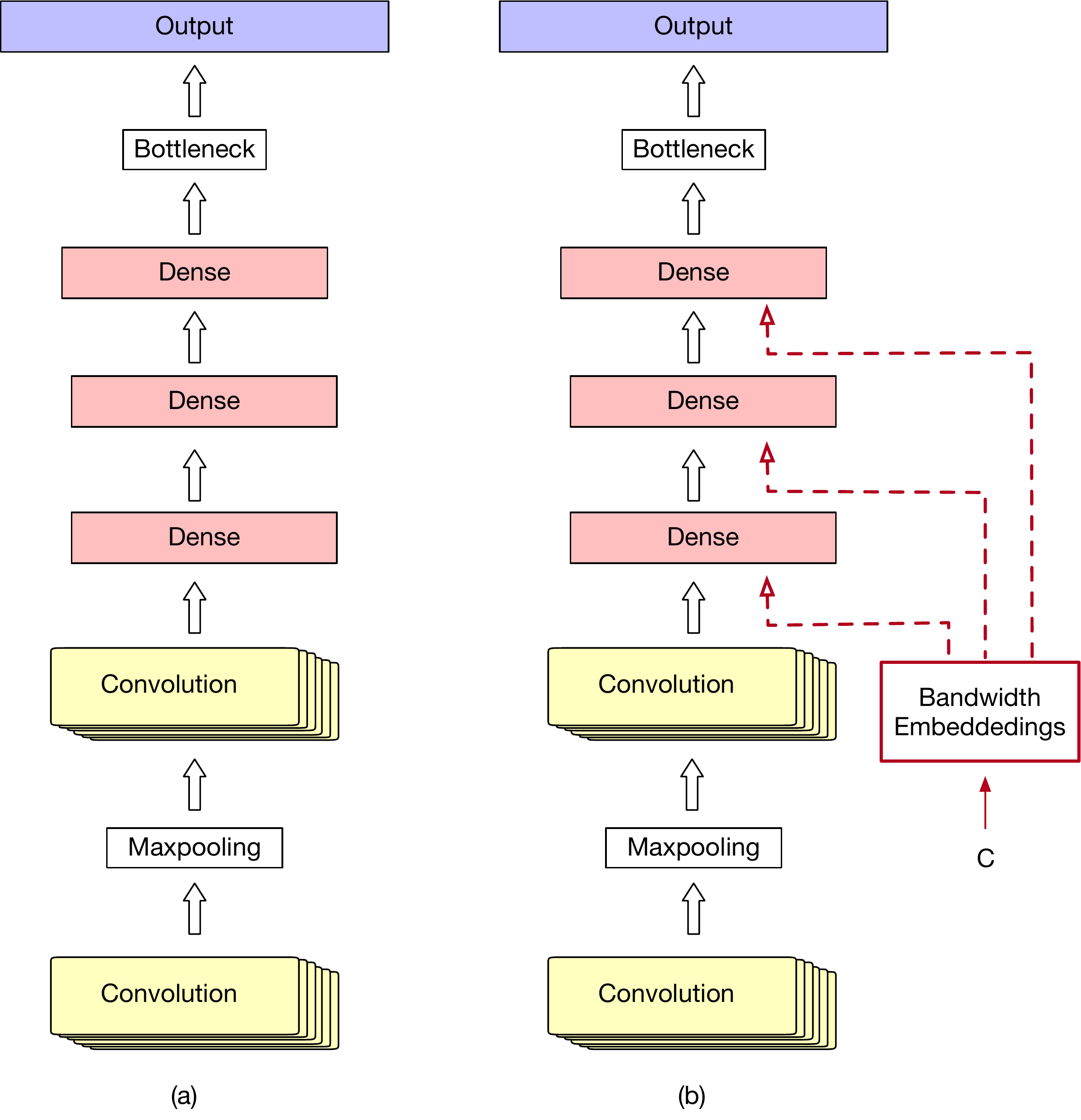}
  \caption{(a) Baseline AM architecture containing two layers of convolution layers, 3 layers of fully connected layers, a liner bottleneck layer and then followed by an output layer, (b) Bandwidth embeddings connected to the dense layers of the baseline architecture, where $c$ represents the type of the speech signal}
  \label{fig:nnet_arch}
\end{figure} 
In this paper, we explore modeling approaches and show that variations in the narrow and wideband speech can be learned and handled via embeddings. Fig.~\ref{fig:nnet_arch}(a) shows the architecture of the baseline AM used in this paper. The model consists of convolutional and dense layers. Convolutional layers are used to reduce the spectral variations in the  features and have shown to perform well for speech recognition~\cite{TASLP:OSSAMA:2014,ICASSP:TARA:2013}. 
%In general, large amounts of training data is required to build DNNs. Hence, it is not very economical to transcribe data for each of the categories. It is not a trivial task to combine both narrow and wideband data to train an DNN-based AM. This is because, narrowband speech do not have any information above 4000 Hz and thus very different as compared to the wideband speech. To overcome this issue, an approach is to alter the conventional feature extraction process such as feature expansion or BWE (a brief description is provided in Section~\ref{sec:intro}). 
Fig.~\ref{fig:nnet_arch}(b), shows the corresponding proposed architecture of the AM which uses an embedding layer connected to all dense layers to handle narrowband and wideband speech jointly. Let weights and bias parameters of a dense layer, $l$, are represented by $W_l$ and $\textbf{b}_l$ respectively. The output  of the dense layer is given as:
\begin{equation}
	\mathbf{o}_l = f(W_{l}\mathbf{o}_{l-1} + \mathbf{b}_l),
\end{equation}
where $f(\cdot)$ is a non-linear activation function. Let $\mathbf{e}^c$ be an $n$ dimensional embedding vector. $c$ is a binary flag distinguishing narrow and wideband data. That is, $c=0$ represents wideband and $c=1$ represents narrowband speech. After incorporating the embedding vector, the equation for $\mathbf{o}_l$ is given as follows:
\begin{eqnarray}
	\mathbf{o}_l &=& f(W_{l}\mathbf{o}_{l-1} + V_{l}\mathbf{e}^c + \mathbf{b}_l) \nonumber \\
	&=& f(W_{l}\mathbf{o}_{l-1} + \mathbf{\hat{b}}_l),
\end{eqnarray}
where $\mathbf{\hat{b}}_l = V_{l}\mathbf{e}^c + \mathbf{b}_l$. $V_l$ is the weight matrix connecting the embedding vector $\mathbf{e}^c$ to the dense layer $l$. In this paper, the bandwidth embeddings is connected to the first dense layer ($l=3$) after two convolutional layers. $V_{l}\mathbf{e}^c$ is referred to as a bias correction term and thus $\mathbf{\hat{b}}_l$ can be referred to as corrected bias. This correction helps the model to differentiate and better process the narrow and wideband data. $\mathbf{e}^{c}$ ($c \in \left\{ 0, 1 \right\}$) is an $n$ dimensional embedding vector and randomly initialized. During training, they are treated as model parameters and are updated during back-propogation. During decoding, the model uses the embedding vector based on the type of input speech signal and is provided by $c$.

\section{Parallel Convolutional Layers}
\label{sec:parallel_conv}
\begin{figure}[t]
  \centering
  \includegraphics[scale=0.35]{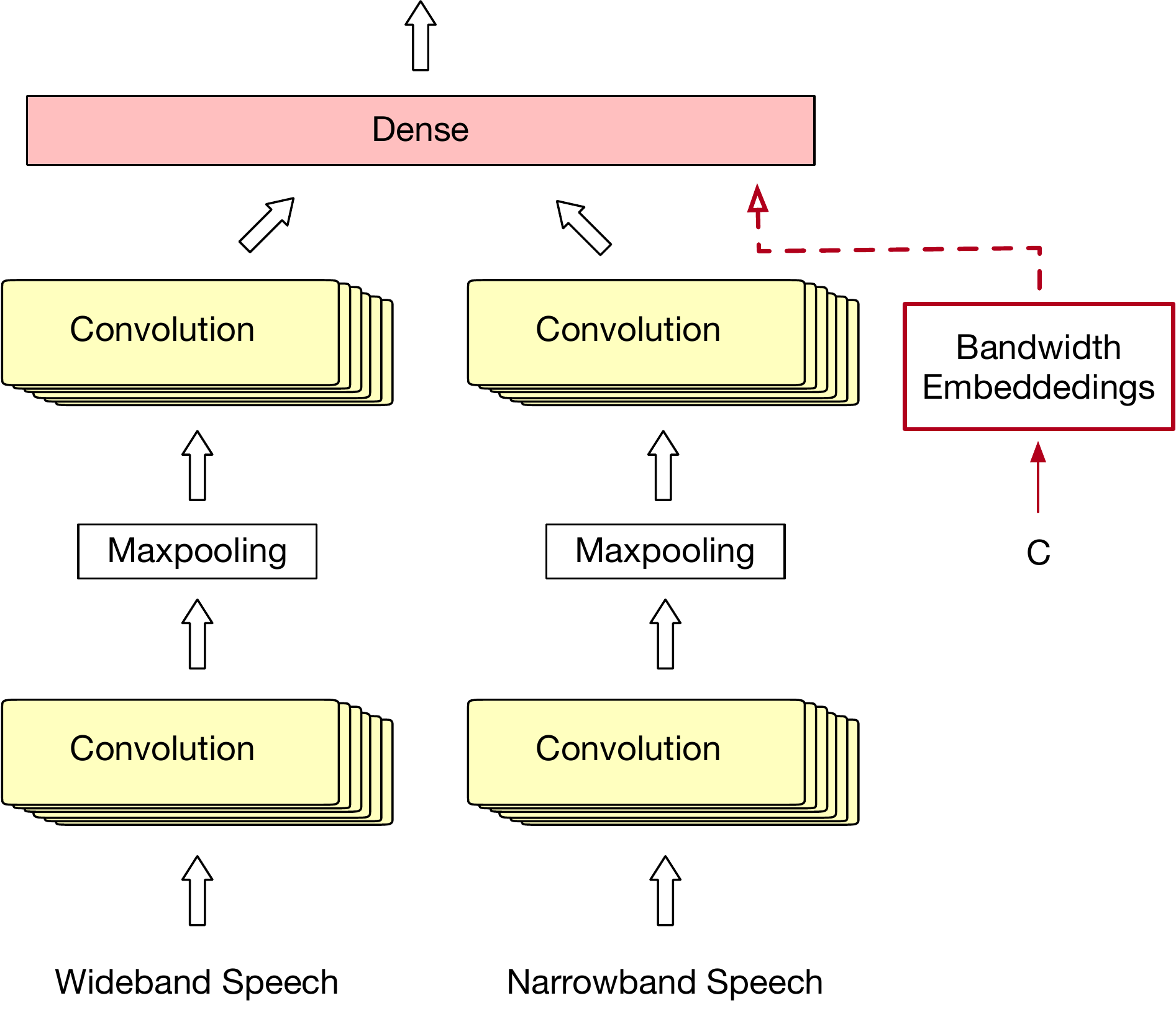}
  \caption{Parallel convolutional layers for narrow and wideband speech.}
  \label{fig:parallel_conv}
\end{figure}

For speech processing, convolutional layers can be considered as a powerful feature processing units. As mentioned earlier, for narrowband speech, the spectral features represent information only from 0-4 kHz and the remaining 4-8 kHz are missing. Hence, use of convolutional layers on features without any prior processing might not be ideal. A simple approach is to upsample the narrowband speech and use the same convolutions for all types of input signals. From a modeling perspective, as an alternative to upsampling, we can use different convolutional layers (as shown in Fig.~\ref{fig:parallel_conv}) for narrow and wideband speech. Filters from these convolutional layers do not share the parameters. We refer to this architecture as parallel convolutional layers. Note that we use shared parameters for the weights connecting the convolutional layers and the dense layer. In Sections ~\ref{sec:database} and ~\ref{sec:evals}, we provide a detailed description of the database used and experimental evaluations to show that embeddings and parallel convolutions can be used to build a single mixed-bandwidth AM which performs well for both the tasks.

\section{Database}
\label{sec:database}
To evaluate our proposed techniques, we use 3400 and 600 hours of wideband (WB) and narrowband (NB) training data. Evaluations are performed on 54 and 4 hours of wideband and narrowband test sets. We report word error rate (WER) to compare the performance of different models. A more detailed description of the data is available in Table~\ref{tab:database}.

\begin{table}[th]
  \caption{Statistics of narrow and wideband training and testing data. Note that \textit{k} used in the numericals represent 1000 units}
  \label{tab:database}
  \centering
  \begin{tabular}{lccc}
    \toprule
    \multirow{2}{*}{\textbf{Data}} & \textbf{Sampling} & \multirow{2}{*}{\textbf{\# Utts.}} & \multirow{2}{*}{\textbf{Hours}}  \\
    & \textbf{Rate (Hz)} & & \\
    \midrule
    \multicolumn{4}{c}{\textbf{Training Data}} \\
    \midrule
    WB & 16k & $2023$k & 3400\\
    NB & 8k & $321$k  & 600\\
    \midrule
    \multicolumn{4}{c}{\textbf{Evaluation Tests}} \\
    \midrule
    WB & 16k & $50$k  & 54.2\\
    NB & 8k & $3.7$k & 4.2\\
    \bottomrule
  \end{tabular}
\end{table}

From Table~\ref{tab:database}, it can be seen that the training data for the wideband speech is much larger than that of the narrowband. In general, such scenarios are common as one often do not have enough training data for each of the task, and wideband speech is more commonly used by many devices including personal assistants these days.
The challenge is to exploit such mixed data for training purposes. %In \cite{INTERSPEECH:XIAODAN:2017}, a similar problem was addressed. That is, narrowband data was limited and thus transfer learning approaches are used to improve the performance of the system. However, in such techniques it is expected to have different models for each of the narrow and wideband speech. 
In Section~\ref{sec:evals}, we provide evaluations and show that the DNNs are powerful and can learn the variations of the narrow and wideband data; thus, avoiding the need for explicit model training for each task.

\section{Evaluations}
\label{sec:evals}

As shown in Fig.~\ref{fig:nnet_arch}, we use a 7 layer deep neural network with 2 convolutional layers, 4 dense layers and followed by an output layer. We use SELU (scaled exponential linear units)~\cite{NIPS:SELU:2017} as activations for all the hidden layers except for the bottleneck layer. The bottleneck layer is a dense layer with linear activations and is often used to reduce the model size~\cite{ICASSP:TARA:2013}. 
We use softmax function as the activations for the output layer. The convolutions used in the first and the second layers consists of 128 filters with kernel sizes of $9\times9$ and $3 \times 4$ respectively. The maxpooling functionality does not have any trainable parameters and hence not considered as layer. The kernel size and the strides used in maxpooling are $1\times3$ and $1\times3$ respectively. The dense, bottleneck and the output layer consists of 1024, 512 and approximately 8000 units. The input to the network are 40 dimensional log mel filter bank features with left and right context of 10.  The model is trained using cross-entropy loss function. We use this architecture in all the evaluations performed in Sections~\ref{subsec:baseline}-\ref{subsec:result_summary}

\subsection{Baseline AMs}
\label{subsec:baseline}

In this section, we present baseline experiments and their results. For training, we use a combination of  wideband and narrowband speech shown in Table~\ref{tab:database}. We built three different AMs: (a) model AM1 built using only wideband speech (WB), (b) AM2 built using only narrowband speech (NB), (c) AM3 built using mixed-bandwidth speech (WB + NB), and (d) AM4 is built using mixed bandwidth data where NB speech is upsampled to 16 kHz. Note that during testing: (a) NB test data is upsampled to 16 kHz when AM1 and AM4 models are used, and (b) WB test data is downsampled to 8 kHz when AM2 model is used. We use sox for resampling the speech data~\cite{sox}.

\begin{table}[th]
  \caption{Evaluating AMs trained using a combination of narrow and wideband data. The word error rates (WER) reported reflect the baseline performance of the ASR systems.}
  \label{tab:baseline_systems}
  \centering
  \begin{tabular}{cccc}
    \toprule
    \multirow{2}{*}{\textbf{Model}} & \textbf{Training} &  \multicolumn{2}{c}{\textbf{WER (\%)}}\\
     & \textbf{Data} & \textbf{WB} & \textbf{NB}\\
    \midrule
    AM1 & WB &  13.1 & ~23.8\tablefootnote{\label{fnote1}Data is upsampled to 16 kHz} \\
    AM2 & NB &  ~22.4\tablefootnote{\label{fnote2}Data is downsampled to 8 kHz} & 21.0 \\
    AM3 & WB + NB & 13.6 & 26.2 \\
    AM4 & ~WB + NB\footref{fnote1} & 13.4 & ~20.9 \footref{fnote1} \\
    \bottomrule
  \end{tabular}
\end{table}

From Table~\ref{tab:baseline_systems}, it can be seen that: (a) for WB test set, AM1 performs better than AM2 and AM3, (b) for NB test set, AM2 performs better than AM1 and AM3, and similar to the performance of AM4. This is because, DNNs tend to perform well in matched conditions. On the other hand, DNNs tend to perform well with increasing amounts of training data. However, in this case, AM3 and AM4 do not reflect such improvement. This is because the spectrum of narrow and wideband speech is different and hence training a model by mixing such data is not trivial. In Section~\ref{subsec:embedded_am}, we show that using bandwidth embeddings we can exploit the use of both narrow and wideband speech for training a single model.

\subsection{Experiments with Bandwidth Embeddings}
\label{subsec:embedded_am}

In this section, we build AMs using an embedding layer connected to the first dense layer as shown in Fig.~\ref{fig:nnet_arch}(b). Two sets of embedding vectors representing the narrow and wideband speech are learned  as part of the AM training and hence simple to use. 

\begin{table}[th]
  \caption{Baseline system performances vs AM trained with embeddings.}
  \label{tab:embedding_system}
  \centering
  \begin{tabular}{lccc}
    \toprule
    \multirow{2}{*}{\textbf{Model}} & \multicolumn{2}{c}{\textbf{WER (\%)}}\\
    &  \textbf{WB} & \textbf{NB}\\
    \midrule
     ~AM1 &  13.1  & ~23.8\footref{fnote1} \\
     ~AM2 & ~22.4\footref{fnote2} & 21.0 \\
     \midrule
     ~AM3 & 13.6  & 26.2 \\
     ~~~~+ Embeddings &  \textbf{12.9}  & \textbf{20.2} \\
      \midrule
     AM4 & 13.4 & ~20.9\footref{fnote1} \\
     ~~~~+ Embeddings & \textbf{13.0} & ~\textbf{18.2}\footref{fnote1} \\
    \bottomrule
  \end{tabular}
\end{table}

In Table~\ref{tab:embedding_system}, we present results where AM3 and AM4 are trained together with the proposed bandwidth embeddings. It can be seen that bandwidth embeddings help to improve the performance of AM3 model with relative improvement of $5\%$ and $23\%$ in word error rate for WB and NB test sets. Also, AM3 + embeddings performs similar to that of AM1 for the WB test set, and performs better than AM2 for NB test set since it can leverage WB data and use more data for training compared to AM2. In Table~\ref{tab:embedding_system}, we see that AM4 + embeddings performs better than all the other systems with a relative improvement in word error rate of $13\%$ as compared to AM2 for the narrowband speech. This is because, to build AM4 models we use the same sampling rate for both types of speech input since narrowband speech is upsampled. Without upsampling of the narrowband speech (e.g. in AM3), the spectral features corresponding to 0-4 kHz do not overlap to that of the features extracted for wideband speech. 
%Using mismatched features for AM training might not be optimal and hence an issue. 
%introduces a mismatch in the spectral information when compared to the wideband speech. That is, features representing the 0-4k Hz in the wideband speech correspond to all the features in narrowband speech.
%Upsampling of 8k Hz data ensures filter banks extracted for narrowband speech overlap with  
These results indicate that a single AM can be used for mixed-bandwidth speech recognition. For an analysis,  we evaluated AM3 model training by varying the embedding vector size from $32$ to $256$ dimensions.

\begin{table}[th]
  \caption{Effect of embedding vector size on the performance of AM3 model.}
  \label{tab:embedding_size}
  \centering
  \begin{tabular}{ccc}
    \toprule
    \multirow{2}{*}{\textbf{Dim.}} & \multicolumn{2}{c}{\textbf{WER (\%)}}\\
     & \textbf{WB} & \textbf{NB}\\
    \midrule
    32 & 13.4 & 21.2\\
    64 & 13.2 & 20.8\\
    \textbf{128} & \textbf{12.9} & \textbf{20.2}\\
    256 & 13.3 & 21.5\\
    \bottomrule
  \end{tabular}
\end{table}

From Table~\ref{tab:embedding_size}, it can be seen that $128$ was the best performing embedding size. Hence, we use 128 dimensional embedding vectors for evaluations in all the experiments.

%In the experiments presented in Table~\ref{tab:embedding_system},  $128$ dimensional embedding vectors are used. In Table~\ref{tab:embedding_dim}, we provide experimental results to analyze the effect of the embedding vector dimension on the speech recognition performance using the AM3 settings. From Table~\ref{tab:embedding_dim}, it can be seen that using embedding vectors of length $128$ performs best for this task; hence it's used in the experiments presented in the rest of the paper.  

%\label{subsec:embedded_dim}
%\begin{table}[th]
%  \caption{WER obtained on WB and NB test sets by varying the dimensions of the embedding vectors. Dimension 0 refers to the baseline AM3 model without any embeddings.}
%  \label{tab:embedding_dim}
%  \centering
%  \begin{tabular}{lcccc}
%    \toprule
%    \multirow{2}{*}{\textbf{Dim.}} & \multicolumn{3}{c}{\textbf{WER (\%)}}\\
%    &  \textbf{WB} & \textbf{NB}\\
%    \midrule
%    0 & 13.6 & 26.2 \\
%     32 & 13.4  & 21.2 \\
%     64 &  13.2  & 20.8 \\
%     128 &  \textbf{12.9} & \textbf{20.2} \\
%     246 &  13.4 & 21.5 \\
%    \bottomrule
%  \end{tabular}
%\end{table}

\subsection{Experiments with Parallel Convolutional Layers}
\label{subsec:parallel_conv}
%Here, we present experimental evaluations using parallel convolutions for narrow and wideband speech. 
%Due the mismatch in narrow and wideband speech features, it may be advantageous to use different sets of convolutions for each type of input speech signal. 
\begin{table}[th]
  \caption{Evaluations performed using parallel convolutions on AM3 and AM4 models.}
  \label{tab:multi_stream_embedding}
  \centering
  \begin{tabular}{lcccc}
    \toprule
    \multirow{2}{*}{\textbf{Model}} & \textbf{Training}& \multicolumn{2}{c}{\textbf{WER (\%)}}\\
    & \textbf{Data} & \textbf{WB} & \textbf{NB}\\
    \midrule
     ~AM3 & WB + NB  & 13.6 & 26.2 \\
     ~~~~+ Embeddings &    &  12.9 & 20.2 \\
     ~~~~+ Parallel Conv.  & & 13.4 & 19.9 \\
     ~~~~+ Embeddings \&   & & \textbf{13.0} & \textbf{19.6} \\
     ~~~~~~~~Parallel Conv.  & & & \\     
     \midrule
     ~AM4 & ~WB + NB\footref{fnote1}  & 13.4 & ~20.9 \\
     ~~~~+ Embeddings &  &  13.0 & ~\textbf{18.2} \\
     ~~~~+ Parallel Conv. & & \textbf{12.7} & ~21.0 \\
     ~~~~+ Embeddings \& &&  13.2 & ~19.9 \\
     ~~~~~~~~Parallel Conv.  & \\
    \bottomrule
  \end{tabular}
\end{table}

In Table~\ref{tab:multi_stream_embedding}, we present results using parallel convolutional layers for AM3 and AM4 setups. It can be seen that using an embedding layer or parallel convolutional layers is improving the performance on either WB or NB or both the test sets. To further improve the system, we explore the use of combining bandwidth embeddings with parallel convolution layers. For the AM3 setup, AM3 + embeddings \& parallel convolutional layers performs the best, and compared to AM3 baseline, it  provides $4\%$ and $25\%$ relative improvement for WB and NB test sets, respectively. Use of parallel convolutional layers for AM3 model training has increased the model size approximately by 200k parameters, which is $1\%$ increase in model parameter size. For the AM4 setup, AM4 + embedding layer performs the best for the NB test set providing $13\%$ relative improvement over the AM4 baseline. Also, experimental results indicate that AM4 setup does not benefit much from the parallel convolution layers for handling the NB test set. This may be due to the fact that since the narrowband data was upsampled to 16 kHz in AM4 training, there is no mismatch in the filter banks used for narrowband and wideband speech; hence there is no need to use separate convolution layers for wideband and narrowband speech data. Whereas in AM3 training, parallel convolution layers help more since narrowband speech is not upsampled, and hence filterbanks used for the wideband and narrowband speech are different. In other words, the parallel convolution layers help to reduce the mismatch between features for the AM3 setup. 
Note that, upsampling of narrowband speech does not provide any new information for the 4k-8k Hz bands. Hence, we believe that AM4 + paralel convolutions is performing well on WB test set by separating convolutional layers for narrowband and wideband speech and possibly reducing the noise that comes from higher frequency of upsampled narrowband speech.
% AM4 model uses more training data than that of AM1 and use of parallel convolutions help to separately process the upsampled narrowband speech with the redundant information. Thus AM4 + parallel convolutions perform better than that of AM1 with a relative improvement of $3\%$ in WER.
 %Hence, differentiating this redundant information could be a reason for AM4 + parallel convolutional layers to perform better than the baseline AM4 model on the WB test set.

%	Also, AM4 + parallel convolutional layers perform well for WB test set. This is because, upsampling of narrowband data is not providing any information for the bands 4k-8k Hz and this redundant information might be the cause of degradation in the baseline AM4 model.
 
% is because there is no mismatch in the features as used in AM3 model training. Thus, from Table~\ref{tab:multi_stream_embedding}, we do not see any improvements when parallel convolutional layers are used with AM4 model.

\subsection{Result Summary}
\label{subsec:result_summary}

In Table~\ref{tab:summary}, we summarize all the evaluations performed from Sections~\ref{subsec:baseline} to ~\ref{subsec:parallel_conv}. AM1 and AM2 are baseline systems which are trained on wideband or narrowband speech respectively. AM3 and AM4 models are trained in combination of bandwidth embeddings and parallel convolutional layers. These models primarily differ in the sampling rate of the train and test data. That is, for AM4 models, narrowband speech was upsampled to 16 kHz, whereas raw narrowband speech was used without any pre-processing in AM3.

%To build the AM4 model, we upsample the narrowband to 16k Hz and this is the major differen

%AM3 model is built on a mixture of 16k and 8k Hz speech data. Thus the best performance is obtained using both embedding and parallel convolutional layers. These parallel convolutions are useful in processing the narrow and wideband speech separately. However, for training AM4 we do not see the same pattern. AM4 training is performed on WT and upsampled NT training data. That is, the complete training data is 16k Hz speech. For AM4, training we consider the use of embedding the best performance system with a WER of $18.2\%$. 

\begin{table}[th]
  \caption{A summary of the performance of different models where: (a) AM1 and AM2 models are the baseline systems, (b) AM3 models are built using a combination of WB and NB data, and (c) AM4 models are built using WB and upsampled NB data.}
  \label{tab:summary}
  \centering
  \begin{tabular}{lccccc}
    \toprule
    \multirow{2}{*}{\textbf{Model}} & \textbf{Training} &  \multicolumn{2}{c}{\textbf{WER (\%)}}\\
    & \textbf{Data} & \textbf{WB} & \textbf{NB}\\
    \midrule
     ~AM1 & WB  & 13.1 & 23.8\footref{fnote1} \\
     ~AM2 & NB & 22.4\footref{fnote2} & 21.0 \\
     \midrule
     ~AM3 & WB + NB  & 13.6 & 26.2 \\
     ~~~~+ Embeddings \&  & &  \textbf{13.0}  & \textbf{19.6} \\
     ~~~~~~~~Parallel Conv.  & \\
     \midrule
     ~AM4 & ~WB + NB\footref{fnote1}  & 13.4 & 20.9 \\
     ~~~~+ Embeddings  & &  \textbf{13.0}  & \textbf{18.2} \\
     %~~~~+ Parallel Conv. & & \textbf{12.7} & 21.0 \\
    \bottomrule
  \end{tabular}
\end{table}

In Table~\ref{tab:summary}, it can be seen that the use of both embeddings and parallel convolutional layers gives the best performance for the model AM3. Compared to AM2 baseline system, we see a relative improvement of $6\%$ in WER for the NB test set, while matching AM1 performance on the WB test set. In AM3, due to the mismatch in the filter bank features, it seems helpful to have separate convolutional layers for narrow and wideband speech. For AM4, compared with the AM2 model, the use of bandwidth embeddings gives a relative improvement of $13\%$ in WER for the NB test set, while matching AM1 performance on the WB test set. AM4 uses upsampled narrowband speech and thus using bandwidth embeddings only seems sufficient. %The models AM3 + embeddings \& parallel convolutional layers and AM4 + embeddings perform similar to that of AM1 for WB test sets.

%To summarize, the best performance systems with the baseline AM1 and AM2 models are: (a) Without re-sampling of the data, we see a relative improvement of $6\%$ for NE test set, and (b) with re-sampling of narrowband speech, we observe a relative improvement of $13\%$ for NE test set using AM4 model. For WE test set, both the models AM3 and AM4 perform similar to that of its baseline AM1 model. Note that, using AM4 model with parallel convolutional layers we see a relative improvement of $3\%$ without any improvement on the NE test set.

\section{Conclusions}
\label{sec:conclusions}
In this paper, we have shown that bandwidth embeddings can be used to build a single model for mixed-bandwidth AM. Further more, we also used different convolutional layers (referred to as parallel convolutional layers) to handle the mismatch between the narrow and wideband speech. Experimental results show that models built using these approaches tend to perform well on narrowband speech without any loss in performance on wideband speech.  For the model trained on wideband and upsampled narrowband speech, using bandwidth embeddings provides a relative improvement of $13\%$ in WER on the narrowband test set while maintaining performance for the wideband speech test. We also showed that using bandwidth embeddings and parallel convolutional layers for 8 kHz and 16 kHz input speech signal has resulted in a relative improvement of $6\%$ in WER on the narrowband test set without requiring upsampling of narrowband speech. 
 %embeddings and parallel convolutions we show a relative improvement of $6\%$ in WER on the narrowband test sets without any need for re-sampling of speech signal.  

\bibliographystyle{IEEEtran}
\bibliography{book,jrnl,conf}
\end{document}